\documentclass[aps,prd,reprint,nofootinbib,superscriptaddress,floatfix,showkeys]{revtex4-1}

\usepackage{amsmath,amssymb}
\usepackage{booktabs}
\usepackage{graphicx}
\usepackage{xcolor}
\usepackage{silence}
\usepackage{hyperref}

\newcommand{\muF}{\mu_{\mathrm f}}
\newcommand{\kD}{\kappa_{\mathrm D}}
\newcommand{\Om}{\Omega}
\newcommand{\Gam}{\Gamma}
\newcommand{\zetaEH}{\zeta_{\mathrm{EH}}}
\newcommand{\revision}[1]{#1}
\newcommand{\additionblue}[1]{#1}
\newcommand{\blueedit}[1]{#1}

\begin{document}

\title{\revision{Model and parameter-path dependence of charged Dirac quasi-bound spectra in Born--Infeld and Euler--Heisenberg-type geometries}}
\author{\additionblue{Bekir Can L{\"u}tf{\"u}o{\u{g}}lu}}
\email{\additionblue{bekir.lutfuoglu@uhk.cz}}
\affiliation{\additionblue{Department of Physics, Faculty of Science, University of Hradec Kr{\'a}lov{\'e}, Rokitansk{\'e}ho 62/26, 500 03 Hradec Kr{\'a}lov{\'e}, Czech Republic}}
\date{\today}

\begin{abstract}
\revision{We study quasi-bound states of a charged massive Dirac field in
the exact electrically charged Einstein--Born--Infeld (BI) geometry and in
a leading-order Euler--Heisenberg-type (EH) approximation.  The two
descriptions agree in the weak-field limit.  We compare families in which
either the mass measured at infinity or the outer-horizon radius is held
fixed, and examine several field charges and levels.  We integrate the
radial equations outward from the horizon and inward from large radius,
then match the solutions.  The resulting frequencies show that the effect
of the nonlinear terms depends on both the electrodynamic theory and the
chosen comparison.  For the fundamental like-charge mode, the small change
in its real frequency has opposite signs in the two theories when the mass
is fixed, whereas it is upward in both when the horizon radius is fixed.
The fundamental mode decays more slowly as the nonlinear term grows along
either family.  The narrowest level, however, decays more
rapidly when the mass is fixed and more slowly when the horizon radius is
fixed.  The spectra of the two theories approach one another as the
nonlinear term is reduced.  We also follow several higher
levels; their decay widths and their response depend on the state and on the
family used for comparison.  The nonlinear coefficient is used as an
effective parameter and is not matched to the microscopic coupling in
\blueedit{quantum electrodynamics (QED)}.}
\end{abstract}

\keywords{\revision{Born--Infeld electrodynamics,
Euler--Heisenberg electrodynamics, charged Dirac fields, quasi-bound states,
spectral response}}

\maketitle

\section{Introduction}

A massive field around a black hole can be trapped outside the horizon even
though the configuration is not exactly stationary in time.  The mass
threshold suppresses propagation to infinity, while gravitational and
electrostatic attraction support a spatially localized cloud.  A small
ingoing component is absorbed by the horizon, giving the mode a finite
complex frequency.  For charged Dirac fields, this problem combines
relativistic spinor dynamics with the competition between gravitational and
Coulomb binding.

\additionblue{Quasi-bound states (QBS)  \cite{DeruelleRuffini:1974} and quasinormal modes (QNMs) \cite{Kokkotas:1999bd, Konoplya:2011qq, Berti:2009kk, Bolokhov:2025rng} solve different radial boundary-value problems. Both are ingoing at the future horizon, but for a massive field below threshold a QBS has an evanescent, decaying solution at large radius, whereas a QNM is selected by an outgoing-wave condition there. QBS therefore describe trapped massive configurations and their leakage into the horizon; QNMs characterize radiative ringing. The spectra are complementary, not interchangeable.}

\additionblue{A qualitative asymmetry in research emphasis accompanies this boundary-condition distinction. Massive-field studies have developed a broad account of mass-dependent ringing and long-lived or quasiresonant spectra for scalar, vector, and fermionic perturbations, across charged, rotating, regular, and modified-gravity backgrounds \cite{Ohashi:2004wr,Konoplya:2002wt,Konoplya:2017tvu,Percival:2020skc,Konoplya:2018qov,Dubinsky:2025bvf,Skvortsova:2024eqi,Lutfuoglu:2025qkt,Bolokhov:2024bke,Lutfuoglu:2025bsf,Lutfuoglu:2025hwh,Lutfuoglu:2026gis,Bolokhov:2026dfg,Skvortsova:2025cah,Skvortsova:2026jtx,Skvortsova:2026unq,Bolokhov:2026dzn}. Other lines of work characterize intermediate and asymptotically late-time tails \cite{Konoplya:2004wg,Konoplya:2006gq,Dubinsky:2024hmn,Bolokhov:2024ixe,Konoplya:2023fmh,Lutfuoglu:2026fpx}, or scattering, transmission, and absorption through grey-body factors and cross sections \cite{Kokkotas:2010zd,Lutfuoglu:2025hjy,Lutfuoglu:2025eik,Lutfuoglu:2025kqp,Lutfuoglu:2026pgn}. Superradiant amplification provides a related route to instability for massive fields around rotating black holes \cite{Konoplya:2008hj}. These observables illuminate radiative response and propagation, but they do not replace the localized-level spectrum selected by a decaying condition at infinity.}

\additionblue{The more focused literature on gravitationally bound or metastable levels began with relativistic energy spectra, quasilevels, and decay widths in Schwarzschild and Kerr settings \cite{DeruelleRuffini:1974,DamourDeruelleRuffini:1976,Ternov:1978,GainaTernov:1988,GainaZaslavskii:1992,Zaslavskii:1990, Konoplya:2026hbz}. Work on rotating black holes then established the role of massive-field bound states in superradiant instability, with charged-field extensions providing a related charged-background setting \cite{Detweiler:1980,Dolan:2007,FuruhashiNambu:2004}. A smaller, related gravitational-atom literature describes long-lived clouds, their excitation, and their radiation, including models with dilatonic or regular interiors \cite{Huang:2022dfx,Dymnikova:2013bna,Corda:2019vuk,Laptev:2006xz,Senjaya:2025ddl,Li:2026gup}. More recent studies extend the spectrum to charged scalar and Proca clouds and charged Dirac modes around Reissner--Nordstr\"om black holes, as well as to Einstein--Gauss--Bonnet and nonlinear-electrodynamic geometries \cite{SampaioEtAl:2014,HuangEtAl:2017,VieiraEtAl:2022,DolanEtAl:2024,KonoplyaStuchlikZhidenko:2026}. The present work adds a comparison of exact Born--Infeld and leading-order Euler--Heisenberg-type backgrounds for the charged-Dirac quasi-bound spectrum.}

\blueedit{Recent work also finds lifetime reversals for neutral and charged scalar clouds around nonminimal Einstein--Yang--Mills black holes \cite{Bolokhov:2026ezi}.}

\revision{The model comparison pursued here is motivated by a useful
overlap between two distinct nonlinear electrodynamics theories.  The
electrically charged Einstein--Born--Infeld solution is known in closed
form, and its weak-field expansion contains the same leading radial
corrections as the Euler--Heisenberg-type geometry used below
\cite{Breton:2003}.  Charged-scalar quasi-bound-state instabilities and
recent analytic instability bounds have been studied in
Einstein--Born--Infeld gravity, while Dirac-field studies in Born--Infeld
backgrounds have focused on quasinormal modes
\cite{WuLu:2024,Hod:2026BornInfeld,MaLi:2020}.  The recent charged-Dirac study in
Einstein--Euler--Heisenberg geometry provides a close point of comparison
\cite{AnarboevEtAl:2026}.  We therefore match the two models at leading
nonlinear order and ask how their finite-coupling Dirac quasi-bound spectra
separate along explicitly specified mass and horizon paths.  To our
knowledge, the massive charged-Dirac quasi-bound spectrum has not previously
been computed on the exact four-dimensional asymptotically flat BI
background.}

\blueedit{These theories are motivated beyond black-hole phenomenology as
well.  At low energies, the Euler--Heisenberg effective action describes
nonlinear QED-vacuum effects, including photon--photon scattering
\cite{HeisenbergEuler:1936}.  In a separate string-theory context,
Born--Infeld-type gauge actions arise in the low-energy world-volume dynamics
of open strings and D-branes \cite{Leigh:1989}.  This connection motivates
the BI model but does not make the Einstein--BI black-hole solution studied
here a string-theory completion.}

Nonlinear electrodynamics changes both the metric and the electric potential
near the black hole.  In the Euler--Heisenberg model, the leading corrections
fall as $r^{-6}$ in the lapse and $r^{-5}$ in the potential.  They therefore
leave the leading $1/r$ Coulomb interaction unchanged but can affect the
near-horizon boundary condition and the absorption width.  The corresponding
static geometry and its thermodynamic properties have been studied in
Refs.~\cite{YajimaTamaki:2001,MagosBreton:2020}.

A numerical comparison of two nonlinear geometries also requires specifying
what is held fixed.  If the ADM mass is fixed, changing the nonlinear
coefficient moves the horizon.  If the horizon radius is fixed, the ADM
mass changes instead.  Since far-zone binding depends on both gravitational
and electrostatic interactions, these two paths need not produce the same
frequency shift.  A statement that a nonlinear correction raises or lowers a level
without this qualification can conflate the local change in the geometry
with a change in the long-range gravitational coupling.

\blueedit{Here we quantify this path dependence by solving the charged-Dirac
radial equations in both the leading-order Euler--Heisenberg-type geometry
and the exact Einstein--Born--Infeld solution of nonlinear Born--Infeld
electrodynamics.  For each model, we compare fixed-mass and
fixed-horizon families that share a Reissner--Nordstr\"om starting point,
vary the field charge, and follow
several fundamental and excited levels.  We report finite-interval
changes in the complex frequencies rather than assigning a universal sign
to the nonlinear response.  This is a controlled spectral study of the
specified geometries and does not claim that the sampled EH coefficients
represent a realistic QED black hole.}

\blueedit{The remainder of the paper is organized as follows. Section~\ref{sec:geometry}
introduces the EH and BI backgrounds and comparison families;
Sec.~\ref{sec:dirac} formulates the charged-Dirac boundary-value problem;
Secs.~\ref{sec:method} and~\ref{sec:numerics} describe the matching procedure,
numerical implementation, and checks; Sec.~\ref{sec:results} presents the
spectra; and Sec.~\ref{sec:discussion} interprets their model and
parameter-path dependence. The Conclusions summarize the principal findings.}

\section{Geometry and comparison families}
\label{sec:geometry}

We use units $G=c=\hbar=1$ and write the static, spherically symmetric
geometry as
\begin{align}
 ds^2&=-f(r)dt^2+\frac{dr^2}{f(r)}+r^2d\Omega^2,
 \label{eq:metric}\\
 A_\mu dx^\mu&=-\Phi(r)dt,
 \label{eq:gauge}
\end{align}
where the gauge is fixed by $\Phi(\infty)=0$.
\blueedit{Here $M$ and $Q$ denote the black hole's Arnowitt--Deser--Misner
(ADM) mass and electric charge, respectively.}
We parameterize the Euler--Heisenberg-type correction by a dimensionless
coefficient $\zetaEH$ and a \blueedit{fixed reference length $M_0$ used to
set the scale for dimensionless parameters}:
\begin{align}
 f(r)&=1-\frac{2M}{r}+\frac{Q^2}{r^2}
 -\frac{\zetaEH M_0^2 Q^4}{20r^6},
 \label{eq:lapse}\\
 \Phi(r)&=\frac{Q}{r}
 -\frac{\zetaEH M_0^2 Q^3}{10r^5}.
 \label{eq:potential}
\end{align}
The factors in Eqs.~\eqref{eq:lapse}--\eqref{eq:potential} define our
normalization of $\zetaEH$.  The coefficient is varied as an effective
parameter and is not matched to the microscopic QED coupling.
\revision{For comparison, we use the exact asymptotically flat electric
Einstein--Born--Infeld solution.  With Born--Infeld scale $\beta$ (of
dimension inverse length), define
$\chi_{\rm BI}(r)\equiv Q^2/(\beta^2r^4)$.  In the same ADM-mass and charge
conventions, the exact solution is}
\begin{align}
 f_{\rm BI}(r)&=1-\frac{2M}{r}
 +\frac{2\beta^2r^2}{3}
 \left(1-\sqrt{1+\chi_{\rm BI}}\right)\notag\\
 &\quad+\frac{4Q^2}{3r^2}
 {}_2F_1\!\left(\frac14,\frac12;\frac54;-\chi_{\rm BI}\right),
 \label{eq:bornInfeldLapse}\\
 \Phi_{\rm BI}(r)&=\frac{Q}{r}
 {}_2F_1\!\left(\frac14,\frac12;\frac54;-\chi_{\rm BI}\right),
 \label{eq:bornInfeldPotential}
\end{align}
where the gauge again satisfies $\Phi_{\rm BI}(\infty)=0$
\cite{Breton:2003}.  Its weak-field expansion is
\begin{align}
 f_{\rm BI}(r)&=1-\frac{2M}{r}+\frac{Q^2}{r^2}\notag\\
 &\quad-\frac{Q^4}{20\beta^2r^6}
 +\frac{Q^6}{72\beta^4r^{10}}\notag\\
 &\quad+O\!\left(\frac{Q^8}{\beta^6r^{14}}\right),\\
 \Phi_{\rm BI}(r)&=\frac{Q}{r}-\frac{Q^3}{10\beta^2r^5}\notag\\
 &\quad+\frac{Q^5}{24\beta^4r^9}
 +O\!\left(\frac{Q^7}{\beta^6r^{13}}\right).
 \label{eq:bornInfeldExpansion}
\end{align}
\revision{Thus the geometries are matched through first order by
$\zetaEH M_0^2=\beta^{-2}$, or equivalently by the dimensionless parameter
$g\equiv\zetaEH=(\beta M_0)^{-2}$.  This matches the leading electric
correction but does not identify the theories: the exact Born--Infeld
solution retains higher powers of $g$, whereas
Eqs.~\eqref{eq:lapse}--\eqref{eq:potential} retain only the leading
correction.  For numerical evaluation we use the following rationalized
form of the BI matter contribution, which remains regular in the Maxwell
limit:}
\begin{equation}
 {
 \begin{aligned}
 h_{\rm BI}(r)&=-\frac{2Q^2}{3r^2[1+\sqrt{1+\chi_{\rm BI}}]}\\
 &\quad+\frac{4Q^2}{3r^2}
 {}_2F_1\!\left(\frac14,\frac12;\frac54;-\chi_{\rm BI}\right).
 \end{aligned}}
 \label{eq:biMatterLapse}
\end{equation}
\additionblue{The function $h_{\rm BI}$ is defined by $f_{\rm BI}(r)=1-2M/r+h_{\rm BI}(r)$.}
\blueedit{When the outer-horizon radius $r_+$ is fixed, the BI ADM mass
follows from $M_{\rm BI}=r_+[1+h_{\rm BI}(r_+)]/2$.}

We set $M_0=1$ and $Q/M_0=0.96$.
The Dirac-field rest mass is \blueedit{$\muF$}, chosen so that
$M_0\muF=0.4$. The field's electric charge is \blueedit{$q$}; we scan
$qM_0=-0.1,0,0.1$. For the matched comparison we sample
$g=0,0.05,0.1,0.25,0.5$ in both models.  At $g=0$ the ADM mass
is $M=M_0$ and the Reissner--Nordstr\"om outer horizon is
\begin{equation}
 r_{+,0}=M_0+\sqrt{M_0^2-Q^2}=1.28M_0.
 \label{eq:rnHorizon}
\end{equation}
The two comparison families are defined as follows:
\begin{enumerate}
 \item In the fixed-mass family, $M=M_0$ and $r_+$ is the largest root of
       $f(r)=0$.
 \item In the fixed-horizon family, $r_+=r_{+,0}$ and the mass is obtained
       from $f(r_+)=0$:
       \begin{equation}
        M(\zetaEH)=\frac12\left[r_++\frac{Q^2}{r_+}
        -\frac{\zetaEH M_0^2Q^4}{20r_+^5}\right].
        \label{eq:massAtFixedHorizon}
       \end{equation}
       \revision{For the exact BI geometry the mass is instead
       $M_{\rm BI}=r_+[1+h_{\rm BI}(r_+)]/2$, with
       $h_{\rm BI}$ given in Eq.~\eqref{eq:biMatterLapse}; in
       the fixed-mass BI family the outer root of
       Eq.~\eqref{eq:bornInfeldLapse} is found numerically.}
\end{enumerate}
Both families therefore represent the same Reissner--Nordstr\"om black hole
at $g=0$.  \revision{At $g=0.5$, the EH truncation gives
$r_+/M_0=1.29324241$ at fixed mass and $M/M_0=0.99691010$ at fixed
horizon; exact BI gives $r_+/M_0=1.29270809$ and $M/M_0=0.99704740$,
respectively.  The outer horizon remains simple throughout the sampled
interval in both models.}

The leading long-distance attractive coupling is
\begin{equation}
 \alpha_{\mathrm C}=M\muF-qQ.
 \label{eq:coulomb}
\end{equation}
\blueedit{This dimensionless coupling combines the gravitational term $M\muF$ with the signed electrostatic term $-qQ$.}
In particular, $\alpha_{\mathrm C}$ is fixed along the fixed-mass path but
varies along the fixed-horizon path.  The nonlinear terms in
Eqs.~\eqref{eq:lapse}--\eqref{eq:potential} do not change the asymptotic
$1/r$ terms.  \revision{The exact BI solution has the same asymptotic
gravitational and electric $1/r$ terms, so this leading-coupling argument
applies to both backgrounds.}

\section{Charged Dirac equation and spectral conditions}
\label{sec:dirac}

Perturbations, separation of variables and analysis of quasinormal modes of the Dirac field have been analyzed in numerous works (see, for instance \cite{Brill:1957fx,Cho:2004wj,Cho:2003qe,Konoplya:2007zx,Kanti:2006ua,Bolokhov:2026kqu,Bolokhov:2026eqf,Malik:2024nhy,Malik:2025erb,Konoplya:2014lha,Lutfuoglu:2026boa}).
The test spinor obeys
\begin{equation}
 \left[i\gamma^\mu\left(\nabla_\mu-iqA_\mu\right)-\muF\right]\psi=0.
 \label{eq:dirac}
\end{equation}
\additionblue{In this normalization, $q$ has dimensions of inverse
length, so $qQ$ and $qM_0$ are dimensionless.}
Using spinor spherical harmonics and a radial rescaling, the angular
eigenvalue is $\kD=\pm(j+1/2)=\pm1,\pm2,\ldots$,
\additionblue{where $j=1/2,3/2,\ldots$ is the total angular momentum.}
With time dependence
$e^{-i\omega t}$, the radial functions satisfy
\begin{align}
 \left(\frac{d}{dr_*}-i\Om(r)\right)R_1
 &=\sqrt{f}\left(\frac{\kD}{r}+i\muF\right)R_2,
 \label{eq:radial1}\\
 \left(\frac{d}{dr_*}+i\Om(r)\right)R_2
 &=\sqrt{f}\left(\frac{\kD}{r}-i\muF\right)R_1,
 \label{eq:radial2}
\end{align}
where $dr_*/dr=f^{-1}$ and $\Om(r)=\omega-q\Phi(r)$.
\revision{A quasi-bound state is a complex-frequency solution that is
ingoing at the future horizon and spatially localized at infinity.  In the
radial normalization of Eqs.~\eqref{eq:radial1}--\eqref{eq:radial2}, the
horizon condition is}
\begin{equation}
 {
 \begin{aligned}
 R_2&\sim e^{-i\Omega_H r_*},\qquad r_*\longrightarrow-\infty,\\
 \frac{R_1}{R_2}&\longrightarrow0,\\
 \Omega_H&\equiv\omega-q\Phi_H,\qquad \Phi_H\equiv\Phi(r_+).
 \end{aligned}}
 \label{eq:horizonBoundary}
\end{equation}
\blueedit{We write the mode frequency as $\omega=\omega_R+i\omega_I$, where
$\omega_R$ and $\omega_I$ are its real and imaginary parts.}
\revision{The asymptotic decay condition is specified by}
\begin{equation}
 k=\sqrt{\muF^2-\omega^2},\qquad \operatorname{Re}k>0.
 \label{eq:decay}
\end{equation}
\revision{The radial spinor then falls as $e^{-kr}$, up to algebraic
Coulomb prefactors.  For the positive-frequency levels considered here,
$0<\omega_R<\muF$.  A QNM instead has the same ingoing horizon condition
but selects the outgoing wave at infinity (with leading phase
$e^{+ik_\infty r_*}$ above the mass threshold, where
$k_\infty=\sqrt{\omega^2-\muF^2}$).  Thus the distinction is the outer
boundary condition: QBS are evanescent and localized, while QNMs are
radiative.}
We report the binding energy $B=\muF-\omega_R$ and the energy width
\begin{equation}
 \Gam=-2\omega_I>0.
 \label{eq:width}
\end{equation}
For this convention, all the computed modes are decaying.  The charged
Dirac field does not exhibit the classical bosonic superradiant amplification
that would provide a growth condition for these modes
\cite{HuangEtAl:2017}.

\section{Two-sided integration and response measures}
\label{sec:method}

We evolve the ratio of the radial components,
\begin{equation}
 Y(r)=\frac{R_1(r)}{R_2(r)},
 \label{eq:ratio}
\end{equation}
which obeys the Riccati equation
\begin{equation}
 \frac{dY}{dr}=\frac{2i\Om}{f}Y
 +\frac{\kD/r+i\muF}{\sqrt f}
 -\frac{\kD/r-i\muF}{\sqrt f}Y^2.
 \label{eq:riccati}
\end{equation}
This avoids integrating the exponentially large overall spinor amplitude.
Writing $x=r-r_+$ and $f=f'_+x+O(x^2)$,
\additionblue{with $f'_+\equiv f'(r_+)$}, the ingoing horizon condition gives
\begin{align}
 Y(x)&=C_H\sqrt{x}+O(x^{3/2}),\\
 C_H&=\frac{\kD/r_+ +i\muF}
 {\sqrt{f'_+}\left[\tfrac12-2i(\omega-q\Phi_H)/f'_+\right]}.
 \label{eq:horizonSeries}
\end{align}
At large radius, the inward-decaying solution has
\begin{align}
 Y(r)&=Y_0+\frac{Y_1}{r}+O(r^{-2}),\\
 Y_0&=\frac{-\omega-ik}{\muF},\\
 Y_1&=-\frac{1}{2k}\left[
 2i(2M\omega-qQ)Y_0+\kD+i\muF M\right.\\
 &\qquad\left.-(\kD-i\muF M)Y_0^2\right].
 \label{eq:infinitySeries}
\end{align}
\revision{The nonlinear corrections in both backgrounds enter beyond
the orders shown because they decay faster than $1/r$.}

We integrate Eq.~\eqref{eq:riccati} outward from $r_++\epsilon$ and inward
from $R_\infty$, and match both ratios at $r_m$.
\additionblue{Here $\epsilon>0$ is the horizon offset, $R_\infty$ is
the outer starting radius, and $r_m$ is the matching radius. The
horizon- and infinity-originating solutions are denoted by $Y_H$ and
$Y_\infty$, respectively.}
The complex residual is
\begin{equation}
 \mathcal E(\omega)=Y_H(r_m;\omega)-Y_\infty(r_m;\omega).
 \label{eq:matchingResidual}
\end{equation}
\additionblue{It is dimensionless because $Y$ is a ratio of radial components.}
The two real equations $\operatorname{Re}\mathcal E=0$ and
$\operatorname{Im}\mathcal E=0$ determine $\omega_R$ and $\omega_I$.
Initial real frequencies use the leading Coulomb estimate
\begin{equation}
 \frac{\omega_R}{\muF}\simeq
 1-\frac{\alpha_{\mathrm C}^2}{2N^2},\qquad
 N=n_r+|\kD|+\Theta(\kD),
 \label{eq:hydrogenic}
\end{equation}
\blueedit{Here $N$ is the principal Coulomb index, $n_r=0,1,2,\ldots$ is
the radial overtone, and $\Theta(\kD)=1$ for positive $\kD$ and zero otherwise.}
The imaginary part is initialized with several decreasing trial magnitudes for the
narrower levels.  At nonzero $\zetaEH$, each mode is continued from its
converged value at the preceding parameter point; no externally tabulated
frequencies are used as root-search seeds.

To compare finite parameter changes, we define the secant responses
\begin{align}
 S_{\omega}&=\frac{M_0[\omega_R(0.5)-\omega_R(0)]}{0.5},\\
 S_{\Gamma}&=\frac{\ln[\Gamma(0.5)/\Gamma(0)]}{0.5}.
 \label{eq:responses}
\end{align}
These are interval-averaged responses, not derivatives at $\zetaEH=0$.
\blueedit{In the tables, $\Delta\omega_R\equiv\omega_R(0.5)-\omega_R(0)$ denotes the real-frequency shift between the endpoint nonlinear couplings.}
The frequencies in the tables are quoted in units of $M_0^{-1}$, including
for the fixed-horizon family where $M$ varies.

\section{Numerical implementation}
\label{sec:numerics}

The boundary-value problem is solved by integrating the ratio equation from
both endpoints rather than by evolving a single spinor from one boundary to
the other.  This avoids the large overall amplitude variation of a bound
spinor and lets the two physical boundary conditions be imposed separately.
The numerical implementation uses the endpoint expansions above as local
initial data, an adaptive high-order integrator in logarithmic radial
coordinates, and a two-real-variable root search for the complex frequency.

\subsection{Endpoint initialization and integration}

Let $\mathcal R(r,Y;\omega)$ denote the right-hand side of
Eq.~\eqref{eq:riccati}.  On the horizon side, we use the offset
$x=r-r_+$ and initialize at $x=\epsilon$ with
$Y=C_H\sqrt{\epsilon}$.  Evolving in $u=\ln x$ gives
\begin{equation}
 \frac{dY}{du}=x\,\mathcal R(r_++x,Y;\omega),
 \qquad \epsilon\leq x\leq r_m-r_+.
 \label{eq:horizonLogIntegration}
\end{equation}
This coordinate resolves the near-horizon region without requiring uniform
steps in $r$ across its many scales.  From infinity, we initialize at
$r=R_\infty$ with $Y=Y_0+Y_1/R_\infty$ and integrate inward in
$v=\ln r$:
\begin{equation}
 \frac{dY}{dv}=r\,\mathcal R(r,Y;\omega),
 \qquad R_\infty\geq r\geq r_m.
 \label{eq:infinityLogIntegration}
\end{equation}
At every trial frequency, we choose the square root
$k=\sqrt{\muF^2-\omega^2}$ so that $\operatorname{Re}k>0$ and the
outer solution decays.
The $Y_0+Y_1/R_\infty$ start retains the first Coulomb correction to the
asymptotic ratio; the omitted terms are suppressed by further inverse
powers of $R_\infty$.

Both complex initial-value problems are integrated with the adaptive
Dormand--Prince method of order eight (DOP853).  For the main survey we use
$r_m=5M_0$, $\epsilon=10^{-10}M_0$, and $R_\infty=1600M_0$. The relative tolerance is
\blueedit{$\mathrm{rtol}=5\times10^{-11}$}, and the absolute tolerance is
$5\times10^{-13}$.
The integrations return the horizon- and infinity-originating ratios at
the same matching radius; only those values enter the residual in
Eq.~\eqref{eq:matchingResidual}.

\subsection{Frequency root search and mode tracking}

We represent the unknown frequency by the real vector
$\boldsymbol{x}=(\omega_R,\omega_I)$ and solve the pair of equations
\begin{equation}
 \boldsymbol{F}(\boldsymbol{x})=
 \begin{pmatrix}
  \operatorname{Re}\mathcal E(\omega_R+i\omega_I)\\
  \operatorname{Im}\mathcal E(\omega_R+i\omega_I)
 \end{pmatrix}
 =\boldsymbol{0}.
 \label{eq:realRootSystem}
\end{equation}
The root is found with a hybrid Powell method, using a step tolerance of
$10^{-10}$ and a maximum of 1600 function evaluations.  The real part of
the initial frequency is the hydrogenic estimate in
Eq.~\eqref{eq:hydrogenic}; for the imaginary part, several negative seed
magnitudes are tried because narrow states can have widths many orders of
magnitude smaller than their binding energies.  A candidate is retained
only if it has $0<\omega_R<\muF$, $\omega_I<0$, and a matching residual
$|\mathcal E|<10^{-8}$.  Thus convergence of the nonlinear solver alone is
not used as the acceptance criterion.

For each state, the starting point at $\zetaEH=0$ is obtained from these
hydrogenic and imaginary-part seeds.  The converged complex root is then
used as the initial guess at the next nonlinear parameter, here
$\zetaEH=0.25$ and $0.5$.  We follow each solution as the nonlinear
parameter varies along its comparison family, rather than using frequencies
reported elsewhere as numerical input.  The mode labels $(n_r,\kD)$ specify the
hydrogenic level used for initialization and continuation; the final
frequency is determined by the two-sided boundary-value condition, not by
the initial approximation.

\revision{For the matched model comparison, we use the same horizon and
infinity conditions, integration limits, root-finding procedure, and mode
labels for the exact BI background.  We continue the BI and EH solutions
independently from the common Reissner--Nordstr\"om solution as the
nonlinear parameter increases.  The analytic BI lapse derivative used to
initialize the horizon expansion is}
\begin{equation}
 f'_{\rm BI}(r)=\frac{1-f_{\rm BI}(r)}{r}
 -\frac{2Q^2}{r^3\left[1+\sqrt{1+Q^2/(\beta^2r^4)}\right]}.
 \label{eq:bornInfeldSlope}
\end{equation}

\subsection{Numerical checks}

We record $|\mathcal E|$ for each root and independently vary the matching
radius, horizon offset, outer starting radius, and integration tolerance.
For the narrow $N=2$, $\kD=-2$ mode, \additionblue{we used the following three parameter choices:}
\begin{align*}
 &(r_m,\epsilon,R_\infty,\mathrm{rtol})
  =(4M_0,10^{-8}M_0,1000M_0,10^{-9}),\\
 &(r_m,\epsilon,R_\infty,\mathrm{rtol})
  =(5M_0,10^{-10}M_0,1600M_0,5\times10^{-11}),\\
 &(r_m,\epsilon,R_\infty,\mathrm{rtol})
  =(7M_0,10^{-11}M_0,2400M_0,2\times10^{-11}).
\end{align*}
Since these quantities are varied together, this comparison tests their
combined effect and does not isolate the error due to each change.  The
maximum residual among the 36 survey frequencies is
$3.03\times10^{-9}$.  Neither a small matching residual nor stability
under these variations is, by itself, a rigorous bound on the frequency
error; together they provide checks on the numerical reliability of the
reported solutions.  The Schwarzschild comparison in
Sec.~\ref{sec:results} supplies a separate check of the radial equations
and boundary implementation.

\section{Results}
\label{sec:results}

\subsection{Dependence on the comparison family}

Table~\ref{tab:main2Protocol} follows the fundamental $\kD=-1$ level for
$qM_0=0.1$.  Its real frequency decreases slightly along the fixed-mass
family, by $4.16\times10^{-6}M_0^{-1}$ between $\zetaEH=0$ and $0.5$.  Along
the fixed-horizon family it increases by $9.44\times10^{-5}M_0^{-1}$.  The
width decreases in both cases, but by different fractions: the endpoint
ratios are $0.99084$ and $0.97053$, respectively.  Thus even at identical
Reissner--Nordstr\"om starting data, the sign of the small real-frequency
shift depends on the family used to vary the nonlinear term.

\begin{table*}[t]
\caption{Fundamental $\kD=-1$ mode at fixed $M_0=1$, $Q/M_0=0.96$, $M_0\muF=0.4$, and $qM_0=0.1$. The two families are defined by holding either the ADM mass or the Reissner--Nordstr\"om outer-horizon radius fixed as the nonlinear parameter is varied. Frequencies and widths are in units of $M_0^{-1}$.}
\label{tab:main2Protocol}
\begin{ruledtabular}
\begin{tabular}{crrrrrr}
Path & $\zetaEH$ & $r_+/M_0$ & $M/M_0$ & $M_0\omega_R$ & $M_0\Gamma$ & $|\mathcal{E}|$ \\
\hline
Fixed $M$ & 0.00 & 1.28000000 & 1.00000000 & 0.3755495982 & 2.20079835e-02 & 2.12e-11 \\
Fixed $M$ & 0.25 & 1.28683056 & 1.00000000 & 0.3755462886 & 2.19069062e-02 & 2.31e-11 \\
Fixed $M$ & 0.50 & 1.29324241 & 1.00000000 & 0.3755454337 & 2.18062982e-02 & 1.61e-11 \\
Fixed $r_+$ & 0.00 & 1.28000000 & 1.00000000 & 0.3755495982 & 2.20079835e-02 & 2.12e-11 \\
Fixed $r_+$ & 0.25 & 1.28000000 & 0.99845505 & 0.3755954870 & 2.16853930e-02 & 3.75e-11 \\
Fixed $r_+$ & 0.50 & 1.28000000 & 0.99691010 & 0.3756439848 & 2.13594800e-02 & 3.35e-11 \\
\end{tabular}
\end{ruledtabular}
\end{table*}

The distinction is not confined to one charge.  Table~\ref{tab:main2Response}
reports the change in the fundamental frequency and width for three values
of $qM_0$ in the EH truncation.  The fixed-mass real-frequency shifts are
negative throughout this charge sample.  At fixed horizon radius, the
like-charge mode shifts upward while the opposite-charge and neutral modes
shift downward.  The opposite-charge width is nearly unchanged along the
fixed-mass path, whereas it decreases along the fixed-horizon path.  These
results caution against assigning a charge-independent sign to the response
without first specifying the geometric quantity held fixed.

\begin{table}[t]
\caption{Finite-interval response of the fundamental $\kD=-1$ mode in the EH geometry. $R_\Gamma\equiv\Gamma(0.5)/\Gamma(0)$; both tabulated quantities are dimensionless.}
\label{tab:main2Response}
\footnotesize
\setlength{\tabcolsep}{3pt}
\begin{ruledtabular}
\begin{tabular}{lcrr}
Path & $qM_0$ & $M_0\Delta\omega_R$ & $R_\Gamma$ \\
\hline
Fixed $M$ & -0.1 & -3.2915280e-04 & 1.0009491 \\
Fixed $M$ & +0.0 & -1.5800331e-04 & 0.9957134 \\
Fixed $M$ & +0.1 & -4.1644795e-06 & 0.9908358 \\
Fixed $r_+$ & -0.1 & -3.5278297e-04 & 0.9943334 \\
Fixed $r_+$ & +0.0 & -1.2503394e-04 & 0.9847257 \\
Fixed $r_+$ & +0.1 & +9.4386543e-05 & 0.9705333 \\
\end{tabular}
\end{ruledtabular}
\end{table}

The reason is visible already in the far-zone coupling.  Along the
fixed-mass family, $\alpha_{\mathrm C}$ is constant, so the leading
hydrogenic energy does not move with $\zetaEH$.  Along the fixed-horizon
family, Eq.~\eqref{eq:massAtFixedHorizon} reduces $M$ as $\zetaEH$ increases,
which changes $\alpha_{\mathrm C}$ even though $Q$ and $\muF$ are held
fixed.  The residual shifts beyond this leading estimate also depend on the
near-horizon potential, the horizon position, and the angular structure of
the spinor.  They cannot be attributed to the metric correction alone.

\subsection{Levels with a common Coulomb index}

The nonrelativistic formula groups several states under the same leading
index $N$.  For $qM_0=0.1$, the states $(n_r,\kD)=(1,-1)$, $(0,+1)$, and
$(0,-2)$ all have $N=2$.  The numerical frequencies and widths are compared
at the Reissner--Nordstr\"om point and at $\zetaEH=0.5$ in
Table~\ref{tab:main2Multiplet}.  Their finite-frequency real parts are
resolved by relativistic and finite-radius effects omitted from
Eq.~\eqref{eq:hydrogenic}.  More strikingly, their widths span roughly four
orders of magnitude: the $\kD=-2$ mode leaks far less efficiently through
the horizon than the $(1,-1)$ level.

\begin{table*}[t]
\caption{Three levels with the same leading nonrelativistic Coulomb index $N=2$ at $qM_0=0.1$. Their finite-frequency splittings include relativistic, curvature, and horizon-boundary effects; their changes with $\zetaEH$ quantify the nonlinear response.}
\label{tab:main2Multiplet}
\begin{ruledtabular}
\begin{tabular}{lcrrrr}
Path & $\zetaEH$ & $(n_r,\kD)$ & $M_0\omega_R$ & $M_0\Gamma$ & $|\mathcal{E}|$ \\
\hline
Fixed $M$ & 0.00 & $(1,-1)$ & 0.3939848387 & 3.93860875e-03 & 1.61e-12 \\
Fixed $M$ & 0.00 & $(0,+1)$ & 0.3942801201 & 2.92789152e-04 & 3.82e-12 \\
Fixed $M$ & 0.00 & $(0,-2)$ & 0.3950500213 & 3.83472166e-07 & 4.91e-10 \\
Fixed $M$ & 0.50 & $(1,-1)$ & 0.3939803553 & 3.91299184e-03 & 3.17e-12 \\
Fixed $M$ & 0.50 & $(0,+1)$ & 0.3942805913 & 2.91755194e-04 & 1.04e-12 \\
Fixed $M$ & 0.50 & $(0,-2)$ & 0.3950500037 & 4.05559064e-07 & 1.42e-11 \\
Fixed $r_+$ & 0.00 & $(1,-1)$ & 0.3939848387 & 3.93860875e-03 & 1.61e-12 \\
Fixed $r_+$ & 0.00 & $(0,+1)$ & 0.3942801201 & 2.92789152e-04 & 3.82e-12 \\
Fixed $r_+$ & 0.00 & $(0,-2)$ & 0.3950500213 & 3.83472166e-07 & 4.91e-10 \\
Fixed $r_+$ & 0.50 & $(1,-1)$ & 0.3940164619 & 3.83535916e-03 & 4.50e-12 \\
Fixed $r_+$ & 0.50 & $(0,+1)$ & 0.3943426895 & 2.74060840e-04 & 8.12e-13 \\
Fixed $r_+$ & 0.50 & $(0,-2)$ & 0.3950949501 & 3.59078970e-07 & 3.11e-11 \\
\end{tabular}
\end{ruledtabular}
\end{table*}

The nonlinear response of the narrowest level changes sign between the two
comparison families.  For $(n_r,\kD)=(0,-2)$, its width grows by about
$5.8\%$ at fixed ADM mass but falls by about $6.4\%$ at fixed horizon
radius.  By contrast, the two other $N=2$ states become slightly narrower
along both paths.  The finite-interval changes in their real frequencies
are small compared with their separation at $\zetaEH=0$, so the ordering of
these three levels is unchanged over the sampled interval.  The result
illustrates why the width should be reported state by state rather than
extrapolated from the fundamental level.

\subsection{Higher excited levels}

We have also followed three higher levels at $qM_0=0.1$:
$(n_r,\kD)=(2,-1)$, $(1,+1)$, and $(1,-2)$, each with Coulomb index
$N=3$.  At the Reissner--Nordstr\"om point their widths range from about
$1.1\times10^{-3}M_0^{-1}$ to $1.6\times10^{-7}M_0^{-1}$.  The first two
levels become narrower along both comparison families.  The narrow
$(1,-2)$ level instead widens slightly at fixed mass and narrows at fixed
horizon radius.  Its real frequency falls by a very small amount at fixed
mass and rises at fixed horizon; the other two levels rise along the
fixed-horizon family, while their fixed-mass shifts are small and of
opposite signs.  The exact BI and EH frequencies remain close.  Table
\ref{tab:main2BIHigherLevels} gives the common Reissner--Nordstr\"om values
and the endpoint frequencies and widths in both theories.

\begin{table*}[t]
\caption{Three higher levels with Coulomb index $N=3$ at $qM_0=0.1$. The Reissner--Nordstr\"om values are common to both families; the EH and BI columns give the endpoint at $g=0.5$. Frequencies and widths are in units of $M_0^{-1}$.}
\label{tab:main2BIHigherLevels}
\begin{ruledtabular}
\footnotesize
\begin{tabular}{crrrrrrr}
Path & $(n_r,\kD)$ & $M_0\omega_R(0)$ & $M_0\Gamma(0)$ & $M_0\omega_R^{\mathrm{EH}}$ & $M_0\Gamma^{\mathrm{EH}}$ & $M_0\omega_R^{\mathrm{BI}}$ & $M_0\Gamma^{\mathrm{BI}}$ \\
\hline
Fixed $M$ & $(2,-1)$ & 0.3974491201 & 1.13753e-03 & 0.3974481492 & 1.13031e-03 & 0.3974483396 & 1.13044e-03 \\
Fixed $M$ & $(1,+1)$ & 0.3975455303 & 1.13308e-04 & 0.3975456842 & 1.12891e-04 & 0.3975456908 & 1.12914e-04 \\
Fixed $M$ & $(1,-2)$ & 0.3977892046 & 1.61232e-07 & 0.3977891978 & 1.70300e-07 & 0.3977891978 & 1.70089e-07 \\
Fixed $r_+$ & $(2,-1)$ & 0.3974491201 & 1.13753e-03 & 0.3974664918 & 1.10695e-03 & 0.3974658819 & 1.10812e-03 \\
Fixed $r_+$ & $(1,+1)$ & 0.3975455303 & 1.13308e-04 & 0.3975709202 & 1.06162e-04 & 0.3975698094 & 1.06479e-04 \\
Fixed $r_+$ & $(1,-2)$ & 0.3977892046 & 1.61232e-07 & 0.3978091270 & 1.50604e-07 & 0.3978082440 & 1.51226e-07 \\
\end{tabular}
\end{ruledtabular}
\end{table*}

\subsection{Exact Born--Infeld versus the EH truncation}

\revision{The exact BI comparison tests which features survive beyond
the shared leading nonlinear term.  At weak coupling, the two spectra
approach one another quadratically: for the like-charge fundamental level,
$|M_0(\omega_R^{\rm BI}-\omega_R^{\rm EH})|$ grows from
$5.39\times10^{-8}$ to $2.11\times10^{-7}$ along the fixed-mass path as
$g$ rises from $0.05$ to $0.1$, and from $7.68\times10^{-9}$ to
$2.92\times10^{-8}$ at fixed horizon.  This is consistent with the
$O(g^2)$ difference expected after matching the $O(g)$ background terms.}

\revision{At $g=0.5$, higher BI orders reverse the sign of the tiny
fixed-mass real-frequency response of the $qM_0=0.1$ fundamental mode:
$M_0\Delta\omega_R=-4.16\times10^{-6}$ in the EH truncation but
$+2.80\times10^{-7}$ in exact BI.  The reversal occurs because the EH
response is already close to zero, not because the two spectra separate
substantially.  At fixed horizon both shifts remain positive and close,
$9.44\times10^{-5}$ and $9.47\times10^{-5}$, respectively.  For this
fundamental level, the width ratios are $0.99084$ (EH) and $0.99111$ (BI)
at fixed mass, and $0.97053$ and $0.97173$ at fixed horizon.}

\begin{table*}[t]
\caption{Like-charge fundamental mode along both comparison paths. The EH and BI coefficients are matched by $g=\zeta_{\mathrm{EH}}$; for BI, $g=(\beta M_0)^{-2}$.}
\label{tab:main2BICurve}
\begin{ruledtabular}
\begin{tabular}{lcrrrrr}
Model & Path & $g$ & $r_+/M_0$ & $M/M_0$ & $M_0\omega_R$ & $M_0\Gamma$ \\
\hline
EH (truncated) & Fixed $M$ & 0.00 & 1.28000000 & 1.00000000 & 0.3755495982 & 2.20080e-02 \\
EH (truncated) & Fixed $M$ & 0.05 & 1.28140284 & 1.00000000 & 0.3755487326 & 2.19878e-02 \\
EH (truncated) & Fixed $M$ & 0.10 & 1.28278672 & 1.00000000 & 0.3755479699 & 2.19675e-02 \\
EH (truncated) & Fixed $M$ & 0.25 & 1.28683056 & 1.00000000 & 0.3755462886 & 2.19069e-02 \\
EH (truncated) & Fixed $M$ & 0.50 & 1.29324241 & 1.00000000 & 0.3755454337 & 2.18063e-02 \\
EH (truncated) & Fixed $r_+$ & 0.00 & 1.28000000 & 1.00000000 & 0.3755495982 & 2.20080e-02 \\
EH (truncated) & Fixed $r_+$ & 0.05 & 1.28000000 & 0.99969101 & 0.3755585777 & 2.19437e-02 \\
EH (truncated) & Fixed $r_+$ & 0.10 & 1.28000000 & 0.99938202 & 0.3755676546 & 2.18794e-02 \\
EH (truncated) & Fixed $r_+$ & 0.25 & 1.28000000 & 0.99845505 & 0.3755954870 & 2.16854e-02 \\
EH (truncated) & Fixed $r_+$ & 0.50 & 1.28000000 & 0.99691010 & 0.3756439848 & 2.13595e-02 \\
BI (exact) & Fixed $M$ & 0.00 & 1.28000000 & 1.00000000 & 0.3755495982 & 2.20080e-02 \\
BI (exact) & Fixed $M$ & 0.05 & 1.28139627 & 1.00000000 & 0.3755487864 & 2.19878e-02 \\
BI (exact) & Fixed $M$ & 0.10 & 1.28276110 & 1.00000000 & 0.3755481809 & 2.19678e-02 \\
BI (exact) & Fixed $M$ & 0.25 & 1.28668147 & 1.00000000 & 0.3755475261 & 2.19084e-02 \\
BI (exact) & Fixed $M$ & 0.50 & 1.29270809 & 1.00000000 & 0.3755498783 & 2.18124e-02 \\
BI (exact) & Fixed $r_+$ & 0.00 & 1.28000000 & 1.00000000 & 0.3755495982 & 2.20080e-02 \\
BI (exact) & Fixed $r_+$ & 0.05 & 1.28000000 & 0.99969247 & 0.3755585854 & 2.19440e-02 \\
BI (exact) & Fixed $r_+$ & 0.10 & 1.28000000 & 0.99938783 & 0.3755676838 & 2.18804e-02 \\
BI (exact) & Fixed $r_+$ & 0.25 & 1.28000000 & 0.99849057 & 0.3755956376 & 2.16920e-02 \\
BI (exact) & Fixed $r_+$ & 0.50 & 1.28000000 & 0.99704740 & 0.3756443468 & 2.13859e-02 \\
\end{tabular}
\end{ruledtabular}
\end{table*}
\revision{The endpoint comparison across charges and levels is given in
Table~\ref{tab:main2BILevels}.  The same-$N$ level ordering is unchanged.
The width differences between the two models are generally below half a
percent, including for the very narrow $(n_r,\kD)=(0,-2)$ state.  Its
relative width response remains qualitatively the same in both models: it
increases by $5.76\%$ (EH) and $5.62\%$ (BI) at fixed mass, but decreases by
$6.36\%$ (EH) and $5.98\%$ (BI) at fixed horizon.  Thus the choice of comparison family
controls the sign of this width response, while the constitutive
model makes a smaller quantitative correction for the parameter range
studied.}

\subsection{Numerical checks}

The radial solver is checked against a massive Dirac frequency in the
Schwarzschild limit.  For $M=1$, $r_+=2$, $M\muF=0.39$, $q=0$, and
$\kD=-1$, it returns
\begin{equation}
 M\omega=0.3630759020-0.0256770637i,
 \label{eq:schwarzschildCheck}
\end{equation}
which agrees with the published value rounded to five decimal places,
$0.36308-0.02568i$ \cite{HuangEtAl:2017}.  At $\zetaEH=0$, the equations
reduce directly to the Reissner--Nordstr\"om case.  The weak-binding
estimate in Eq.~\eqref{eq:hydrogenic} is used for initialization and as a
check on the leading long-range attraction; it is not expected to reproduce
the finite-coupling level shifts or widths.

The Reissner--Nordstr\"om limit also permits comparison with
published charged-Dirac results.  For $Q/M_0=0.96$ and $M_0\muF=0.4$,
the fundamental $\kD=-1$ mode changes from
$(M_0\omega_R,M_0\Gam)=(0.34835,0.08271)$ at $qM_0=-0.1$ to
$(0.37555,0.02201)$ at $qM_0=+0.1$.  Thus, as the field charge changes
from opposite to like sign with the black-hole charge, the real frequency
rises and the width falls.  The same trend is found in the Reissner--Nordstr\"om
charge study of Huang et al.~\cite{HuangEtAl:2017}.  Their parameters differ
from those used here, so the agreement concerns the direction of the change,
not the individual frequency values.

We also vary the matching point, horizon offset, outer starting radius, and
integration tolerance for the particularly narrow $N=2$, $\kD=-2$ mode.
\additionblue{Across these runs, its displayed real frequency is unchanged at
the quoted precision, and the matching residuals range from
$4.3\times10^{-11}$ to $1.4\times10^{-8}$. These residuals measure the
quality of the matching but do not give a rigorous bound on the eigenfrequency
error.}

\revision{For the matched survey, the largest residual among the 180
EH and exact-BI roots is $2.47\times10^{-9}$.  The BI geometry checks recover
the Maxwell limit and reproduce the matched first-order lapse and potential
to absolute differences below $3\times10^{-14}$ at $g=10^{-5}$; the analytic
lapse slope agrees with a finite-difference check to $7.8\times10^{-12}$.
\additionblue{The BI fundamental and narrow $\kD=-2$ frequencies are
likewise stable under these variations; the matching residuals range from
$8.9\times10^{-12}$ to $1.5\times10^{-8}$.}}

The added narrow $N=3$, $\kD=-2$ mode was checked by varying
the matching radius, horizon offset, outer radius, and integration tolerance.
Its real frequency varies by less than $2\times10^{-12}$ across the three
integrations, and the matching residuals range
from $2.0\times10^{-13}$ to $7.4\times10^{-11}$.

\begin{table*}[t]
\caption{Exact Born--Infeld and truncated Euler--Heisenberg frequencies at matched leading coupling $g=\zeta_{\mathrm{EH}}=0.5$. The frequencies and widths are in units of $M_0^{-1}$.}
\label{tab:main2BILevels}
\begin{ruledtabular}
\begin{tabular}{crrrrrrr}
Path & $qM_0$ & $(n_r,\kD)$ & $M_0\omega_R^{\mathrm{EH}}$ & $M_0\omega_R^{\mathrm{BI}}$ & $M_0\Gamma^{\mathrm{EH}}$ & $M_0\Gamma^{\mathrm{BI}}$ & $\Gamma_{\mathrm{BI}}/\Gamma_{\mathrm{EH}}$ \\
\hline
Fixed $M$ & -0.1 & $(0,-1)$ & 0.3480246782 & 0.3480315933 & 8.27893e-02 & 8.27554e-02 & 0.999592 \\
Fixed $M$ & +0.0 & $(0,-1)$ & 0.3617982049 & 0.3618078853 & 4.81363e-02 & 4.81286e-02 & 0.999841 \\
Fixed $M$ & +0.1 & $(0,-1)$ & 0.3755454337 & 0.3755498783 & 2.18063e-02 & 2.18124e-02 & 1.000281 \\
Fixed $M$ & +0.1 & $(1,-1)$ & 0.3939803553 & 0.3939810601 & 3.91299e-03 & 3.91339e-03 & 1.000102 \\
Fixed $M$ & +0.1 & $(0,+1)$ & 0.3942805913 & 0.3942806057 & 2.91755e-04 & 2.91820e-04 & 1.000222 \\
Fixed $M$ & +0.1 & $(0,-2)$ & 0.3950500037 & 0.3950500036 & 4.05559e-07 & 4.05036e-07 & 0.998711 \\
Fixed $r_+$ & -0.1 & $(0,-1)$ & 0.3480010479 & 0.3480091905 & 8.22421e-02 & 8.22301e-02 & 0.999854 \\
Fixed $r_+$ & +0.0 & $(0,-1)$ & 0.3618311743 & 0.3618400305 & 4.76051e-02 & 4.76203e-02 & 1.000319 \\
Fixed $r_+$ & +0.1 & $(0,-1)$ & 0.3756439848 & 0.3756443468 & 2.13595e-02 & 2.13859e-02 & 1.001236 \\
Fixed $r_+$ & +0.1 & $(1,-1)$ & 0.3940164619 & 0.3940156131 & 3.83536e-03 & 3.83922e-03 & 1.001008 \\
Fixed $r_+$ & +0.1 & $(0,+1)$ & 0.3943426895 & 0.3943399560 & 2.74061e-04 & 2.74897e-04 & 1.003051 \\
Fixed $r_+$ & +0.1 & $(0,-2)$ & 0.3950949501 & 0.3950929587 & 3.59079e-07 & 3.60523e-07 & 1.004020 \\
\end{tabular}
\end{ruledtabular}
\end{table*}

\section{Discussion}
\label{sec:discussion}

\revision{The main result is methodological as well as spectral: a
nonlinear-coupling response is a property of a specified family of black-hole
solutions and constitutive model, not of a metric coefficient considered in
isolation.  Both paths begin at the same $(M,Q,r_+)$ in the Maxwell limit.
The fixed-mass path preserves the leading gravitational Coulomb term while
moving the horizon; the fixed-horizon path changes the mass and therefore
the leading coupling $M\muF-qQ$.  In exact BI and in the EH truncation, the
fixed-horizon like-charge fundamental shifts upward.  At fixed mass, their
finite-interval shifts have opposite signs, but the difference is small in
absolute terms and reflects a near-cancellation in the EH response.}

\revision{The width response remains state dependent in both models.
The narrow $(0,-2)$ level widens at fixed ADM mass but narrows at fixed
horizon radius; the two descriptions differ by less than half a percent in
its endpoint width at $g=0.5$.  The same-$N$ levels retain their ordering,
although their widths span several orders of magnitude.  These results show
that the comparison path controls the qualitative response in this sample,
while higher Born--Infeld orders give smaller but measurable quantitative
changes.}

\revision{The scope is limited to static, electrically charged,
asymptotically flat backgrounds and a test Dirac field.  We do not include
field backreaction, discharge, or environmental screening.  The EH geometry
is a leading-order effective truncation and its coefficient is not matched
to microscopic QED; the exact BI solution is a distinct nonlinear model,
not a completion of the EH theory.  Results are given for $Q/M_0=0.96$ and
$0\leq g\leq0.5$.  Broader charge ratios, weaker binding, and more
near-extremal paths remain to be explored.}
\section{Conclusions}

Quasinormal-mode studies of charged black holes span a broad range of
geometries and perturbing fields
\cite{Kokkotas:1988fm,Onozawa:1996ba,Onozawa:1995vu,Konoplya:2002wt,Berti:2003ud,Konoplya:2002ky,Kodama:2003kk,Kodama:2007ph,Konoplya:2008rq,Konoplya:2013sba,Konoplya:2007jv,Davey:2023fin,Fernando:2004pc,Panotopoulos:2019qjk,Panotopoulos:2019tyg,Churilova:2019sah,Konoplya:2001ji,Fernando:2012yw,Flachi:2012nv,Toshmatov:2015wga,Breton:2016mqh,Macedo:2016yyo,MahdavianYekta:2019pol,Lopez:2022uie,Bolokhov:2023ruj}. By comparison, massive quasi-bound states form a more focused literature, with charged scalar and Proca clouds and charged Dirac modes studied around Reissner--Nordstr\"om black holes \cite{SampaioEtAl:2014,HuangEtAl:2017}.

\revision{We have compared charged Dirac quasi-bound spectra in the exact
four-dimensional Einstein--Born--Infeld geometry with those in its matched
leading-order EH-type truncation.  The common weak-field terms produce
quadratically converging spectra as $g\to0$, while higher BI orders reverse
the sign of the nearly vanishing fixed-mass real-frequency response of the
like-charge fundamental state.  The fixed-horizon response keeps the same
sign in both models, and the narrowest sampled state's width changes in
opposite directions between the two parameter paths.  Three
additional excited levels show the same dependence on the state and on the
chosen family, with the narrowest again widening at fixed mass and narrowing
at fixed horizon.  The comparison
therefore separates constitutive-model corrections from parameter-path
effects.}

\section*{Note added}

\textit{Note added (17 October 2026).}  A related study of charged Dirac
quasi-bound states in Einstein--Euler--Heisenberg geometry was posted on
arXiv on 12 October 2026 \cite{AnarboevEtAl:2026}.  In addition to the
fixed-mass versus fixed-horizon comparison, the present work compares the
leading-order EH-type geometry with the exact Einstein--Born--Infeld
background at matched weak-field coupling.

\begin{acknowledgments}
The author acknowledges the use of ChatGPT (OpenAI) for assistance with
LaTeX editing and language polishing. The author takes responsibility for
the final manuscript.
\end{acknowledgments}

\bibliographystyle{apsrev4-1}
\raggedright
\bibliography{references}

\end{document}